\documentclass[twocolumn,preprintnumbers,amsmath,amssymb,superscriptaddress]{revtex4}
\usepackage{amsmath,amsfonts,amssymb}
\usepackage{color}

\usepackage{amsmath,amsfonts,amssymb}
\usepackage[english]{babel}
\usepackage[T1]{fontenc}
\usepackage{color}
\usepackage{float}
\usepackage{verbatim}
\usepackage{graphicx}
\usepackage{bm}
\usepackage{mathtools}
\usepackage{stmaryrd}
\usepackage{anyfontsize}
\usepackage{changepage}

\usepackage{natbib}
\definecolor{darkblue}{rgb}{0,0,0.6}
\definecolor{darkcyan}{rgb}{0.1,0.3,0.4}
\definecolor{darkgreen}{rgb}{0,0.4,0}
\definecolor{darkred}{rgb}{0.6,0,0}

\newcommand{\beginsupplement}{%
        \clearpage
        \setcounter{table}{0}
        \renewcommand{\thetable}{S\arabic{table}}%
        \setcounter{figure}{0}
        \renewcommand{\thefigure}{S\arabic{figure}}%
     }

\newcommand{\ct}{${\rm C}_3~$}

\newcommand{\cf}{${\rm C}_4~$}

\begin{document}

\title{Caching in or falling back at the Sevilleta}

\author{Justin D. Yeakel} \affiliation{School of Natural Sciences, University
  of California, Merced, Merced, CA 95340, USA}

\author{Uttam Bhat} \affiliation{School of Natural Sciences, University
  of California, Merced, Merced, CA 95340, USA}

\author{Seth D. Newsome} \affiliation{Department of Biology, University of New Mexico, Albuquerque, NM 87131, USA}

% \title{Caching in on the Sevilleta: how consumer body size, resource availability, and seasonal uncertainty impacts the foraging strategies of desert rodents}

% This version of the LaTeX template was last updated on
% February 15, 2019.

% Articles
% Major articles contain new data or new theory or new analysis of existing data. Papers proposing a novel method should address how the method advances understanding of a general conceptual issue, illustrate the application of the method to data, and discuss the potential for broader use of the approach.
% 
% The American Naturalist's policy is that papers should be as long as they need to be to make their case well, but the preference is for manuscripts that are approximately 5500 words or fewer of text, not including the literature cited, and have no more than six tables and/or figures. Additional material can appear in the expanded online edition. Such material can include appendixes, tables, and figures as well as electronic enhancements such as video, sound, and data files (see details below). Because each article must stand on its own merits, we do not accept paired articles.

%%%%%%%%%%%%%%%%%%%%%
% Authorship
%%%%%%%%%%%%%%%%%%%%%
% Please remove authorship information while your paper is under review,
% unless you wish to waive your anonymity under double-blind review. You
% will need to add this information back in to your final files after
% acceptance.
% 
% \author{Justin D. Yeakel$^{1,2,\ast}$ \\ 
% Uttam Bhat$^{1,2}$ \\ 
% Seth D. Newsome$^{3}$}

% \date{}

\begin{abstract}

Foraging in uncertain environments requires balancing the risks associated with finding alternative resources against potential gains. In aridland environments characterized by extreme variation in the amount and seasonal timing of primary production, consumer communities must weigh the risks associated with foraging for preferred seeds that can be cached against fallback foods of low nutritional quality (e.g., leaves) that must be consumed immediately. Here we explore the influence of resource-scarcity, body size, and seasonal uncertainty on the expected foraging behaviors of caching rodents in the northern Chihuahaun Desert by integrating these elements with a Stochastic Dynamic Program (SDP) to determine fitness-maximizing foraging strategies. We demonstrate that resource-limited environments promote dependence on fallback foods, reducing the likelihood of starvation while increasing future risk exposure. Our results point to a qualitative difference in the use of fallback foods and the fitness benefits of caching at the threshold body size of $50$g. Above this threshold the fitness benefits are greater for consumers with smaller caches, affirming empirical observations of cache use among rodents in such dynamic environments. This suggests that larger-bodied consumers with larger caches may be less sensitive to the future uncertainties in monsoonal onset predicted by global climate scenarios.
\end{abstract}

\maketitle
% 
% \noindent{} 1. University of California, Merced, Merced, California 95348;
% 
% \noindent{} 2. Santa Fe Institute, Santa Fe, New Mexico;
% 
% \noindent{} 3. University of New Mexico, Albuquerque, New Mexico.
% 
% \noindent{} $\ast$ Corresponding author; e-mail: jyeakel@ucmerced.edu.
% 
% 
% \bigskip
% 
% \textit{Manuscript elements}: Figure~1, figure~2, figure~3, figure~4, figure~5, figure~6, table~1, table~2, online appendices~A and B (including figure~A1, figure~A2, figure~A3, figure~A4). All figures are to print in color.
% 
% \bigskip
% 
% \textit{Keywords}: Foraging, caching, fallback foods, Sevilleta, Stochastic Dynamic Programming
% 
% \bigskip
% 
% \textit{Manuscript type}: Article. %Or e-article, note, e-note, natural history miscellany, e-natural history miscellany, comment, reply, invited symposium, or countdown to 150.
% 
% \bigskip
% 
% \noindent{\footnotesize Prepared using the suggested \LaTeX{} template for \textit{Am.\ Nat.}}
% 
% %\linenumbers{}
% %\modulolinenumbers[3]
% 
% \newpage{}

% 
% \newpage{}
% 
% \linenumbers 
% \modulolinenumbers[2]
% 

\section*{Introduction}

To survive, consumers must take risks as they seek rewards.
The strategies they employ must account for both the immediate costs of today, as well as the uncertainties of tomorrow. %, on into the future to the end of their lives.
This entails weighing the costs and benefits of foraging in a temporally and spatially stochastic environment, conditioned on the energetic state of the individual \cite{Mangel1986}.
While the rewards are energetic and ultimately reproductive, the risks include exposure to predation and the potential that resources are not found such that foraging effort is wasted.

The potential fitness gains provided by specific resources must be weighed in terms of their availability \cite{Caraco1980}, energy density \cite{Emlen1966}, macronutrient content and stoichiometry \cite{SternerElser2002}, and the mechanical and chemical costs associated with processing and digestion \cite{Dominy2008,Nersesian2011,Yeakel2013,Lucas2014,MachovskyCapuska2016}.
% Critical use times and the importance of fallback foods  \emph{fall back}
During times of plenty the quality of easily found resources may be prioritized, whereas during times of scarcity resource availability may be expected to play a larger role.
% Importantly, both the mean density of a resource, as well as the variability in its mean density, or patchiness, is expected to impact foraging decisions \cite{Caraco1980,Bhat2019}.
As such, resources can be categorized in terms of preference, where preferred foods are those targeted when all resources are available, and non-preferred fallback foods are those consumed when preferred foods are less available.

While food preference is a well-understood concept, often less appreciated is the importance of those resources targeted during critical-use times when preferred high-quality resources are limiting \cite{Wrangham1998,Laden2005}.
During these periods, consumers generally \emph{fall back} on resources that are otherwise underutilized, which may be due to differences in energy density, abundance, or the direct (e.g. predation) or indirect risks (e.g. competition) associated with acquisition or assimilation.
Such foods are sometimes referred to as keystone resources \cite{Leighton1983}, however this term is problematic because it has connotations with community stability and the assumption that a keystone resource's out-sized role varies inversely with its abundance \cite{Constantino2009}.
To avoid this confusion, we refer to these resources as \emph{fallback} foods, a term oft-used in primate ecology and paleoanthropology to describe those foods ``of low preference and high importance" \cite{Marshall2009,Lambert2015}.
While fallback resources may be utilized only rarely, they are vital for ensuring survival during hard times.\\

\noindent \emph{``Where he can't save a cent and his whole life is spent / Just waitin' for the hard times to go" \\ -Dave Alvin (Interstate City, 1996)}\\ \\
To `save' for a terrestrial consumer means to accumulate energy either endogenously as fat in the form of adipose tissue, or exogenously in a cache \cite{Smith1984}.
Because small mammals can save proportionately smaller amounts of fat on their bodies than larger mammals, they operate closer to the starvation threshold \cite{Lindstedt1985,Dunbrack1993,Yeakel2018}, meaning that hard times are never far away.
The limited storage afforded to smaller mammals means that they are more sensitive to spatio-temporal fluctuations in resource availability.
Caching food resources can buffer against this foraging uncertainty, and is a particularly common behavior employed by small mammals and passerine birds \cite{Smith1984,Dunbrack1993}.
While many of the advantages of caching have been explored previously \cite{McNamara1990,Vanderwall1990,Lucas1991,Brodin1997,Gerber2004,Mangel1988}, how this behavior advantages consumers of different body sizes in highly seasonal environments, particularly when seasonal transitions are uncertain, is not well understood.

Desert ecosystems often support diverse and dynamic small mammal communities in spite of low and unpredictable resource availability \cite{Fox2011}.
These communities exemplify how resource-limited ecosystems can support consumers with a diverse range of life-history modes and functional traits associated with resource procurement.
In the arid Sonoran and Chihuahuan Deserts of the American Southwest, for example, Heteromyid rodents are food-hoarding (caching) granivores that range in body size from $\sim 5-150$ g and have long gestation times and small litter sizes, but they typically out-compete other species for high-quality seeds.
In particular, larger species such as kangaroo rats (\emph{Dipodomys} spp.) use both scatter and larder-hoarding strategies to store food in caches that can persist across seasons and even years to provide reliable sources of food during periods of resource scarcity \cite{Schroder1979,Vanderwall1990}.
Smaller heteromyids such as the silky pocket mouse (\emph{Perognathus flavus}) have highly plastic diets that limit competitive overlap with co-occurring species \cite{Noble2019}.
In contrast, Cricetids such as deer mice (\emph{Peromyscus} spp.) and grasshopper mice (\emph{Onychomys} spp.) do not generally cache food and thus can only store energy endogenously, but often extract resources from multiple trophic levels. 
With shorter gestation times and larger litter sizes, Cricetids typically have higher reproductive potential than sympatric Heteromyids \cite{Hoffmeister1986}.

The highly seasonal and inter-annual climate variability in desert ecosystems provides an ideal system to quantify the effects of resources on the functional and community ecology of a diverse small mammal community. 
In the northern Chihuahuan Desert of central New Mexico, precipitation is bimodal, with, on average, ~60\% of annual rainfall being delivered by the summer monsoon from Jul–Oct. 
Monthly averages for more unpredictable winter and spring precipitation are lower than the monsoon, while the driest and hottest period of the year is typically in May–Jun.
Another attribute of this system is inter-annual variability in both the (annual or seasonal) amount of precipitation and the timing of the transitions between seasons, which impacts consumer foraging strategies of desert consumers \cite{Orr2015}. %(Orr et al. 2015, Noble et al. 2019). 
Overall, environmental variation is a common attribute of precipitation regimes in arid ecosystems, and it is one reason why such environments have served as the backdrop for field-based experiments examining the influence of abiotic factors such as precipitation and temperature on plant \cite{McDowell2008} and consumer communities \cite{Meserve2003,Chesson2004,Thibault2004,Kelt2011}. 

These two distinct periods of annual precipitation produce resources of differing quantity and quality that can be traced through the consumer community with carbon isotope analysis of primary consumer tissues and the plant resources they utilize. 
Highly unpredictable winter/spring rains fuel a spring period of \ct primary productivity, namely annual forbs and perennial shrubs. 
Later in the summer, a second more reliable period of monsoonal precipitation drives the production of \cf grasses and limited \ct growth. 
\ct and \cf plants vary in their nutritional quality, energy content, and persistence in the environment. 
The leaves of \ct plants are more nutritious with higher nitrogen and digestible carbohydrate contents than \cf grasses \cite{Caswell1973,Caswell1975,Caswell1976,Barbehenn2004,Barbehenn2004b}. 
% \ct annuals also produce larger seeds (Reichman 1976, Harper 1977, Davidson et al. 1985, Samson et al. 1992). 
In contrast, the leaves of \cf grasses are harder for consumers to process and digest, but are more resistant to decomposition \cite{Vanderbilt2008} and may serve as a fallback food for rodents during periods of resource scarcity. 

Regional climate models predict rapidly increasing air temperatures, significant decreases of 15–20\% in winter precipitation, increased inter-annual variability in the strength and onset of the summer monsoon, and higher drought risk \cite{Gutzler2011,Seager2007,Cook2015}. 
These directional shifts in abiotic conditions could push some flora and fauna beyond their physiological or ecological limits for local occupancy, but to date most studies on this topic have focused on strict physiological tolerances to water limitation or temperature \cite{McDowell2008,Breshears2009,McKechnie2010}. 
Rodent consumers, whose lifespans are generally on the order of $\sim 1-3$ years, must adopt foraging strategies to compensate and prepare for this increased environmental variation that will likely impact the phenology of resource availability. 
To what extent different foraging strategies may be best-suited to absorb these changes is therefore of central importance for understanding how small mammal communities in the arid American Southwest will be influenced by future environmental changes.

Here we explore the influence of resource-scarcity, body size, and seasonal uncertainty on the expected foraging behaviors of caching rodents in the northern Chihuahuan Desert at the Sevilleta National Wildlife Refuge.
We integrate these elements using a Stochastic Dynamic Program (SDP) to determine fitness-maximizing foraging strategies, where a strategy is defined by the extent to which different resource functional groups are targeted within a heterogeneous and seasonal landscape.
Our approach offers four important insights.
First, we show that environments characterized by resource scarcity promote consumer dependence on fallback foods that are not cacheable (leaves), increasing risk exposure and limiting the ability of a consumer to absorb the effects of future hardship.
Second, we observe that smaller consumers generally rely on fallback foods to a greater extent, which limits their ability to maintain large caches and buffer against future uncertainties in resource procurement.
Third, we find that while uncertainty in seasonal transitions lowers consumer fitness across the board, maintaining a larger cache can compensate for these fitness detriments, though the extent of compensation depends strongly on both consumer body size and resource availability.
Because the onset of the monsoon season is expected to become less predictable as climate change progresses, the fitness advantage provided to caching consumers may have large implications for the future of the Sevilleta rodent communities in the American Southwest and other stochastic aridland ecosystems.
Finally, our results point to a qualitative difference in the use of fallback foods and the fitness benefits of caching at the threshold body size of $\sim 50$ g.
Above this threshold the fitness benefits are greater for consumers with smaller caches, affirming empirical observations of cache use among rodents at the Sevilleta.

\begin{figure}
\centering
\includegraphics[width=0.5\textwidth]{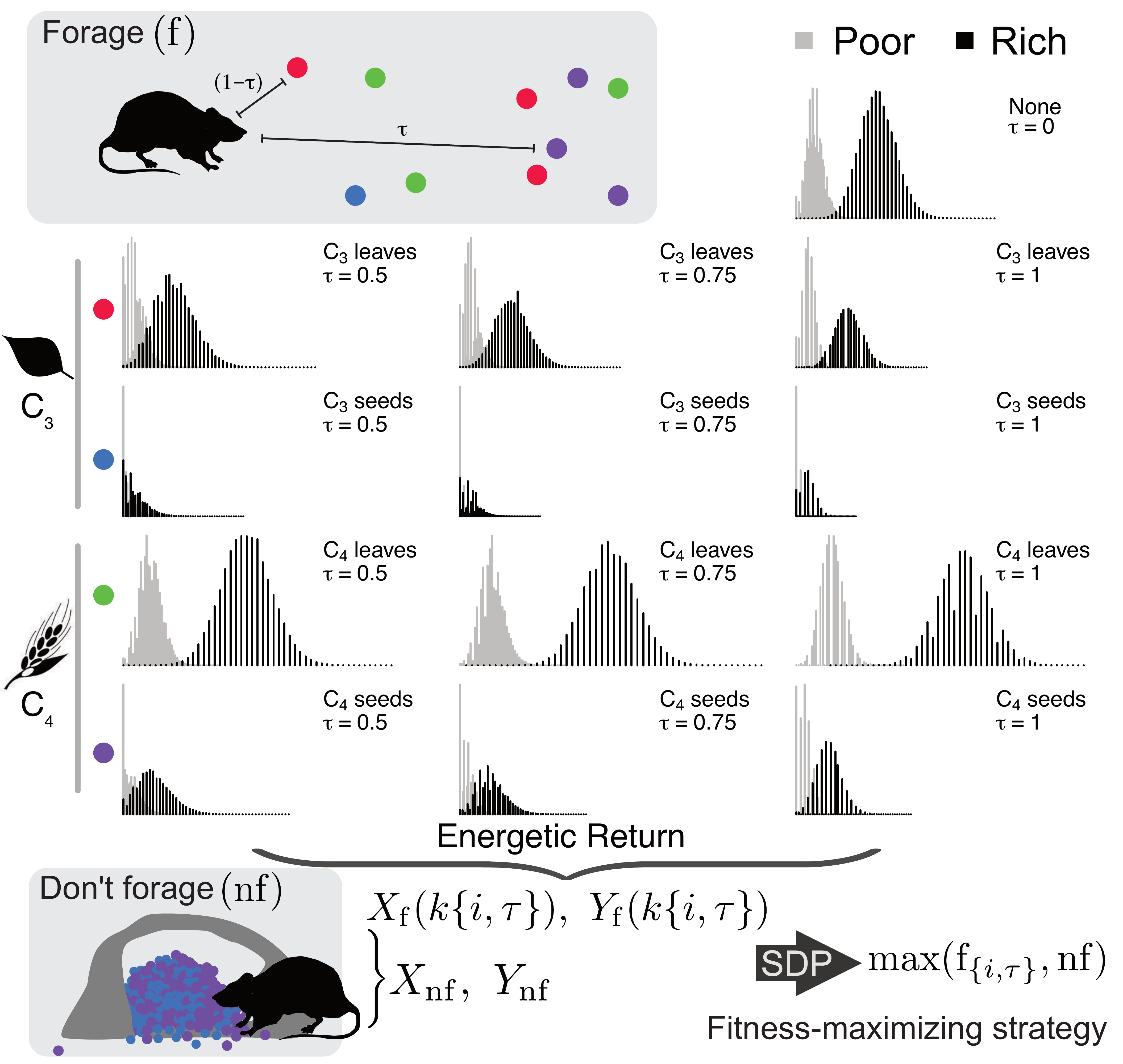}
\caption{
An illustration of the model used to calculate fitness-maximizing foraging strategies.
Foraging strategies are defined by targeting different resources $i$ or by not foraging and relying instead on cached reserves.
A consumer targeting resource $i$ approaches that resource with probability $\tau$ regardless of distance, and the nearest resource with probability $(1-\tau)$.
The probability of potential energetic gains in season $s$ across this strategy-space $k_s\{i,\tau\}$ are numerically determined.
Shown here are energetic return probability distributions for the monsoon season in both poor (low $\rho$; gray) and rich (high $\rho$; black) states.
The consumer may also choose to not forage (nf), whereupon it relies on cached reserves.
We then examine how all possible strategies $({\rm f}_{\{i,\tau\}},{\rm nf})$ change lifetime fitness by altering the consumer's fat stores $X$ and cache reserves $Y$.
The fitness-maximizing foraging strategy is determined by the rules of SDP.
}
\label{fig:model}
\end{figure}

\section*{Methods}
We use a Stochastic Dynamic Program (SDP) to quantify the fitness-maximizing foraging strategies of rodent consumers as a function of \emph{i}) endogenous (on-board) energetic reserves stored as body fat $X=x$, \emph{ii}) exogenous (off-board) energetic reserves stored in a cache $Y = y$, and \emph{iii}) the day $d$ in season $s$ over a year defined by a pre-, full-, post-monsoonal seasonal transition.
We maintain this notation, upper case for random variables and lower case for specific values, for all stochastic quantities.
First we will describe how we simulate consumer-resource interactions within a particular foraging bout across heterogeneous resource landscapes.
The results from this simulation will be used to numerically estimate a probability distribution describing  energetic returns for a consumer implementing a specific foraging strategy.
Second, we will show how the energetic returns that result from a particular foraging strategy contribute to the dynamics of the consumer's energetic state, defined by the amount of fat and cache stores accumulated by the consumer over the $d_{rm max}$ days and $s_{\rm max}$ seasons of the year.
Third, we will describe how these dynamics are integrated into a fitness equation, which will allow us to determine the fitness-maximizing foraging strategy for a consumer as a function of its energetic state $(X=x,Y=y)$ and time of year $(d,s)$.
See fig. \ref{fig:model} for a conceptual illustration of the modeling framework.

\subsection*{Energetic returns within a foraging bout}
In our framework, consumers forage in an environment where there are four resource functional groups: \ct leaves, \ct seeds, \cf leaves, and \cf seeds, where a resource $i$ has a unique mean density $\mu_i$ characteristic of the pre-/post-, and full-monsoon seasons.
Resource mean densities ($\text{g~m}^{-2}$) were estimated from seasonal transects as part of the Sevilleta LTER monitoring effort (table 1).

Rodent consumers are central-place foragers with relatively small home-ranges that are spatially constrained by competition.
Because an individual consumer can cover the same ground multiple times within a foraging bout, resource availability is assumed to scale with home-range area, such that larger home-ranges promote increased resource availability and, by extension, higher effective densities and encounter rates.
In contrast, the number of resources intercepted by a forager unconstrained by a central-place would be expected to scale with velocity rather than home-range area. %and resource density to match that of the spatial distribution.

To consider central-place consumers of different body sizes foraging over home-ranges of different areas, we set the mean encounter rate as $m_i = \rho \mu_i M^{3/4}$ for resource $i$, where the unitless $\rho$ scales environmental richness for all potential resources.
Lower values of $\rho$ correspond to poor environmental conditions, such that all resources in each season are less common -- though their relative availability does not change -- whereas higher values of of $\rho$ correspond to rich environmental conditions.
Each resource $i$ is distributed according to a mean encounter rate $m_i$ and dispersion $\alpha_i$, where lower values of $\alpha$ correspond to patchier distributions.
Because we are considering resources that are expected to have similar spatial clustering, we set $\alpha = 10$ for \ct and \cf leaves and seeds.
We show in the Supplementary Materials that alternative values of $\alpha$, as well as the scaling of home-range area with body size, do not have a significant influence on our results (appendix A).

We estimated energetic return distributions for each foraging strategy.
% where a preferred resource is targeted with probability $\tau$, and resource closest in space is targeted with probably $(1-\tau)$.
A foraging strategy is defined by the targeting of a preferred resource $i$ with probability $\tau$ or the closest resource $j$ with probability $(1-\tau$).
% To estimate the distribution of energetic returns within a foraging bout, we employed a numerical simulation where the consumer repeatedly observes foods according to their spatial distributions, and either \emph{i}) acquires the nearest resource, or \emph{ii}) acquires a preferred, or targeted, resource regardless of distance.
The distance from the consumer to a resource $i$ is drawn from an exponential distribution with mean $1/\Lambda_i$, where the random variable $\Lambda_i \sim {\rm Gamma}(\alpha_i,m_i/\alpha_i)$.
A chosen resource is approached by the consumer with body mass $M$ by a velocity $v$ ($\propto M^{0.21}$), after which the consumer moves on to the next resource based on the next set of independently drawn distances (see fig. \ref{fig:model}).
Metabolic costs during the foraging bout follow the field metabolic rate, whereas metabolic costs are assumed to follow the basal metabolic rate for the remaining time in the day (see table 2).
This foraging process repeats until the temporal window of the bout has closed.
As resources obtained during a foraging bout are assorted, they are added to fat reserves $X$ and/or to the cache $Y$, where we record the proportional contribution of each resource to the total kJ return as $\pi_i$, and in particular those that are cacheable as $\pi_{\rm cache} = \pi_{\rm C_3~seeds} + \pi_{\rm C_4~seeds}$.
Importantly, the potential kJ return of a bout in season $s$, $K_s$, is proportional to the energy density $g_i$ of each found resource $i$ (table 1), such that $K_s = N\sum_i \pi_i g_i$, where $N=n$ are the g of resources gathered in a bout.
By simulating $10^6$ independent replicate bouts, we calculate the probability $p(K_s=k_s)$ of gathering a potential energetic return of $k_s$ kJ for a given foraging strategy in season $s$.\\

\subsection*{Foraging strategies}
The complete set of alternative foraging strategies that we explore includes 1) not foraging and relying instead on cached reserves, 2) foraging without preference for specific resources, and 3) targeting each of the four resource functional groups with increasing preference (fig. \ref{fig:model}).
Fitness is evaluated across different foraging strategies, where resources $i$ are targeted with weights $\tau$; if a resource $i$ is targeted with weight $\tau$, the consumer approaches its preferred resource with probability $\tau$ regardless of distance; with probability $(1-\tau)$, the consumer approaches the nearest resource.
We consider 3 targeting options for each of the 4 resource groups, where targeting weights range from partial targeting ($\tau = 0.5,~0.75$) to complete targeting ($\tau = 1$) on each of the resources.
If no resources are targeted, the consumer approaches and consumes/gathers the closest resource, and its final cumulative returns will mirror the environmental resource distribution.
If a resource is targeted at $\tau=1$, the consumer ignores all resources except that which it is targeting.
Given the targeting weights $\tau = (0.5,0.75,1.0)$, the universe of foraging behaviors includes 14 options: stay home and consume cache, forage without targeting, and forage according to the 3 targeting weights for each of the 4 resources (fig. \ref{fig:model}).\\

\subsection*{Consumer state dynamics}
We assumed a consumer of mass $M$ to have an on-board energetic storage equivalent to its allometrically-determined fat mass in addition to 10\% of its muscle mass, assuming an energy density of $20~\text{kJ~g}^{-1}$ \cite{Stryer1995,Hou2008} (see table 2 for allometric relationships).
We scaled the units of the consumer's fat reserves such that the maximum storage $x_{\rm max}=20$, whereas the critical value at which starvation occurs is $x_{\rm crit} = 1$.
In contrast, the maximum size of the cache is set to last a consumer half a season (50 days), and, unlike fat stores, does not involve a critical value resulting in mortality.

When an organism chooses not to forage (denoted by the subscript `nf') and instead relies on its cache for replenishment,
\begin{align}
  X_{\rm nf}(d+1)&= X(d) - c_{\rm nf} + \epsilon_{\rm cache}{\rm min}[x_{\rm stm}, Y(d)] \nonumber \\
  Y_{\rm nf}(d+1) &= Y(d)(1 - \delta) - {\rm min}[x_{\rm stm},Y(d)]
\end{align}
where $x_{\rm stm}$ is the limit (in kJ) of the consumer's stomach, $\delta$ is the decay rate of the cache, $\epsilon_{\rm cache}$ is the digestibility of seeds, and the energetic cost $c_{\rm nf} = b_{\rm basal}t_{\rm day}$ given that $t_{\rm day}$ is the number of hours in a day.
If the consumer stays home, the energetic cost is relatively low, and it fills its stomach with the digestible portion of its cache.
The cache is depleted by the amount eaten by the consumer as well as per-period decay.

A consumer foraging (denoted by the subscript `f') with a targeting weight $\tau$ on resource $i$, obtains a potential energetic return $k_s\{i,\tau\}$ with probability $p(K_s=k_s\{i,\tau\})$.
Given this potential return, the fat and cache stores change as
\begin{widetext}
\begin{align}
  X_{\rm f}(k_s\{i,\tau\},d+1) &= X(d) - c_{\rm f} + \sum_{j=1}^{n=4} \epsilon_j \pi_j{\rm min}[x_{\rm stm},k_s\{i,\tau\}],  \nonumber \\
  Y_{\rm f}(k_s\{i,\tau\},d+1) &= Y(d)(1 - \delta) + {\rm min}[x_{\rm chk},\pi_{\rm cache}(\underbrace{k_s\{i,\tau\}-{\rm min}[x_{\rm stm},k_s\{i,\tau\}]}_{\rm remainder})]
\end{align}
\end{widetext}
where the energetic cost $c_{\rm f} = b_{\rm field}t_{\rm bout} + b_{\rm basal}(t_{\rm day}-t_{\rm bout})$ given that $t_{\rm bout}$ is the number of hours in a foraging bout, and $x_{\rm chk}$ is the kJ limit of the consumer's cheek pouch.
In words, if the consumer forages it fills its stomach with its energetic returns and saves the cacheable portion of the remainder, limited by the storage capacity of its cheek pouches. %, which we assume is the same as the storage capacity of its stomach.
While the found resources fills the consumer's stomach irrespective of digestibility, only the digestible portion can be added to fat reserves.
Moreover, only the proportion of the energetic return attributable to seeds is cacheable ($\pi_{\rm cache}$), and the cache grows by the amount the consumer returns in its cheek after it fills its stomach. \\

\subsection*{Fitness-maximizing foraging strategies} 
Fitness can be interpreted as the probability of survival during a non-breeding interval of length $d_{\rm max} s_{\rm max}$ where $d_{\rm max}=100$ is the length of each season (in days), and $s_{\rm max}=3$ is the number of seasons in the year.
We define the terminal fitness function for the last day/season of the year for a consumer, which is assumed to increase symmetrically with both fat and cache reserves, such that
\begin{align}
  \Phi(x,y,d=d_{\rm max},s=s_{\rm max}) = 
  \begin{cases}
    \frac{x + \epsilon_{\rm cache} y}{x_{\rm max} + \epsilon_{\rm cache} y_{\rm max}} & \text{if}\ x > x_{\rm crit}\\
    0 & \text{if}\ x=x_{\rm crit}.
  \end{cases}
\end{align}
For times previous to the terminal time, we define the fitness function
\begin{widetext}
\begin{equation}
  F(x,y,d,s) = {\rm max}~{\rm E}\left\{\Phi(X(d_{\rm max},s_{\rm max}),Y(d_{\rm max},s_{\rm max})|X=x,Y=y)\right\},
  \label{eq:F1}
\end{equation}
\end{widetext}
where the maximization selects the foraging behavior that maximizes fitness given fat and cache reserves during day $d$ of season $s$.
For time periods prior the terminal time, an organism must both survive and select the fitness maximizing foraging strategy given the stochasticity of energetic gains.
If the daily mortality probability is $m_{\rm nf}$ for a consumer that stays home and $m_{\rm f}$ if the consumer forages (where we assume that $m_{\rm nf} < m_{\rm f}$), then $F(x,y,d,s)$ satisfies the equations of SDP such that
\begin{widetext}
\begin{align}
  F(x,y,d,s) = &{\rm max}\Bigg[\underbrace{(1-m_{\rm nf})F\{x_{\rm nf}, y_{\rm nf}, d+1, s\}}_{\rm stay~home}, \nonumber \\
  & \left. \underbrace{\underset{(i,\tau)}{\rm max}\left[\overbrace{(1-m_{\rm f})\sum_{k=k_{\rm min}}^{k_{\rm max}}p(k_s\{i,\tau\})F\{x_{\rm f}(k_s\{i,\tau\}), y_{\rm f}(k_s\{i,\tau\}), d+1, s\}}^{\Omega_{i,\tau}(x_{\rm f},y_{\rm f},d+1,s)}\right]}_{\text{target resource } i \text{ with weight }\tau} \right]
  \label{eq:SDP}
\end{align}
\end{widetext}
where seasons are coupled, such that the consumer's fitness on the last day of season $s$ is a function of the consumer's fitness on the first day of season $s+1$.
In this framework, seasonal transitions are deterministic and immediately result in changes to resource distributions.

To implement uncertainty in seasonal transitions, we assume that there is an increased probability that a monsoonal environment will transition to a pre-/post- monsoonal environment, or that the pre-/post- monsoonal environment will transition to a monsoonal environment as $d\rightarrow d_{\rm max}$.
If $s^\prime$ represents the alternative season, we assume that the probability of transition $p(s \rightarrow s^\prime|d,\sigma)$ increases with each day to a maximum value of $p=1/2$ at $d=d_{\rm max}$.
As such, the uncertainty in transitions between seasons is more severe with larger $\sigma$, and deterministic if $\sigma \ll 1$.
With seasonal transition uncertainty, the fitness equation becomes
\begin{widetext}
\begin{align}
  F(x,y,d,s|\sigma) = &{\rm max}\Bigg[(1-m_{\rm nf})F\{x_{\rm nf}, y_{\rm nf}, d+1, s\}, \\ \nonumber
  & \underset{(i,\tau)}{\rm max}\Big[\Big(1-p(s \rightarrow s^\prime|d,\sigma)\Big)\Omega_{i,\tau}(x_{\rm f},y_{\rm f},d+1,s)+ p(s \rightarrow s^\prime|d,\sigma)\Omega_{i,\tau}(x_{\rm f},y_{\rm f},d+1,s^\prime)\Big] \Bigg],
  \label{eq:F3}
\end{align}
\end{widetext}
where we use the $\Omega$ notation defined in Eq. \ref{eq:SDP}.\\

\section*{Results \& Discussion}

\subsection*{When to Cache and When to Fallback}

\begin{figure}
\centering
\includegraphics[width=0.5\textwidth]{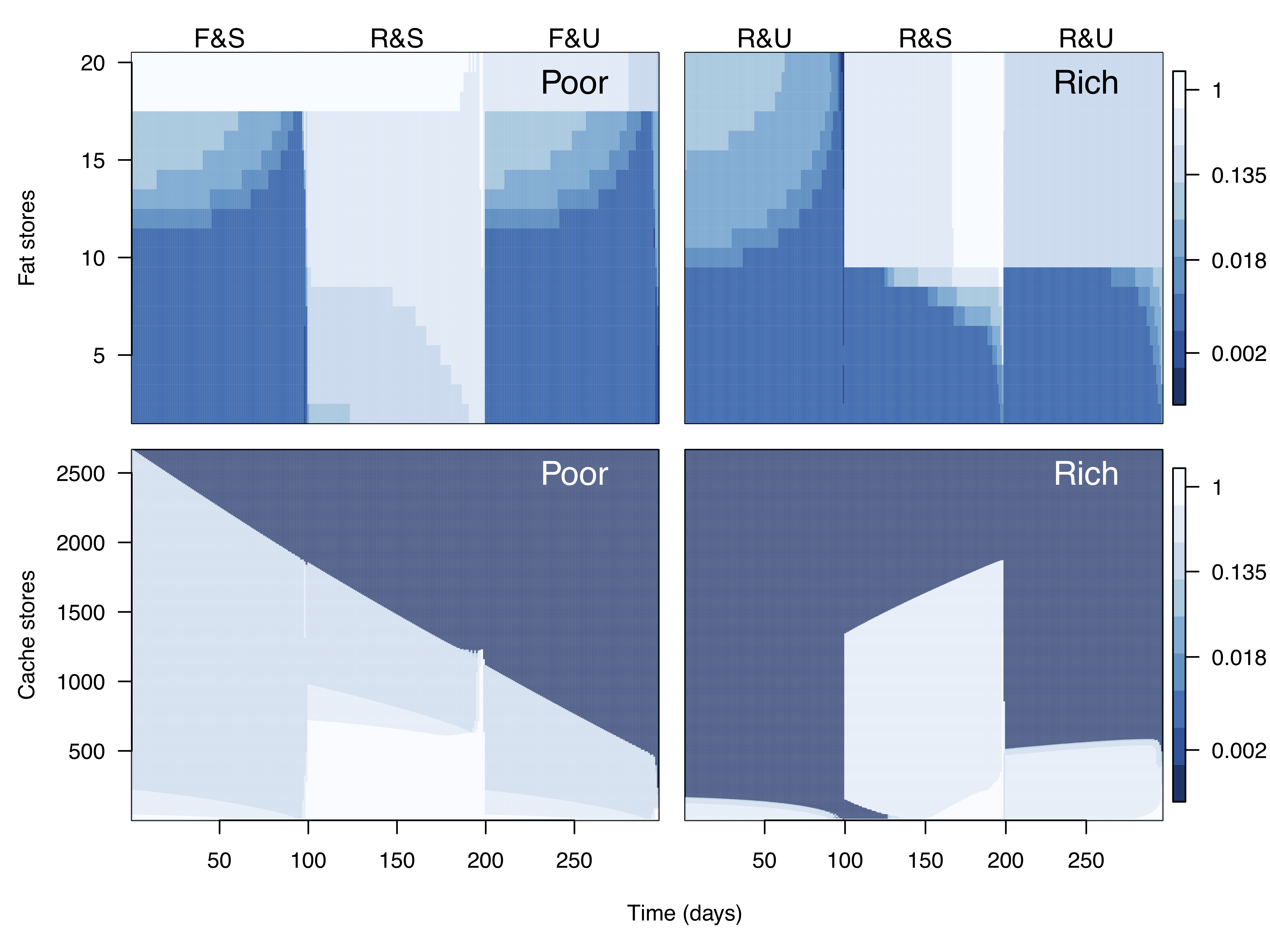}
\caption{
The proportion of states that result in active foraging versus not foraging and relying on cached reserves.
Values range from 100\% states that result in actively foraging (white) to 0\% (dark blue).
0\% active foraging implies 100\% relying on cached reserves.
To observe the dependence of foraging/not foraging strategies on consumer fat ($X$) and cache ($Y$) states as well as environmental richness, values are
(top-left) averaged across cache states in poor environments,
(bottom-left) averaged across fat states in poor environments,
(top-right) averaged across cache states in rich environments, and
(bottom-right) averaged across fat states in rich environments.
Note that the color scale is logged to emphasize smaller differences in consumer strategies.
}
\label{fig:forage}
\end{figure}

We first assessed general consumer behavioral trends over a pre- to post-monsoonal seasonal transition.
This seasonal pattern is a simplification of the spring (pre-monsoon), summer (full-monsoon), fall (post-monsoon) transition characterizing the main pulse in primary productivity in the Sonoran and Chihuahuan Desert each year.
Behavioral trends were assessed by calculating the proportion of consumer states resulting in active foraging or relying on cached resources, averaged over both the consumer's fat stores and cache reserves independently (fig. \ref{fig:forage}).
Assuming deterministic transitions between seasons, we find that consumers employ four primary foraging strategies depending on season and environmental condition.
We define these strategies by combinations of two sets of alternative foraging modes: 
\emph{i}) \emph{fallback} versus \emph{replenish}: whether foraged foods consist of fallback resources (F), or those that can be used to replenish the cache (R), and
\emph{ii}) \emph{use} versus \emph{save}: whether cached resources are used (U) or saved (S) against future hardship.
We assessed when these strategies were implemented across the pre-, full-, and post-monsoon seasonal transition in both resource-poor (low $\rho$) and resource-rich (high $\rho$) environments.

\begin{figure}
\centering
\includegraphics[width=0.5\textwidth]{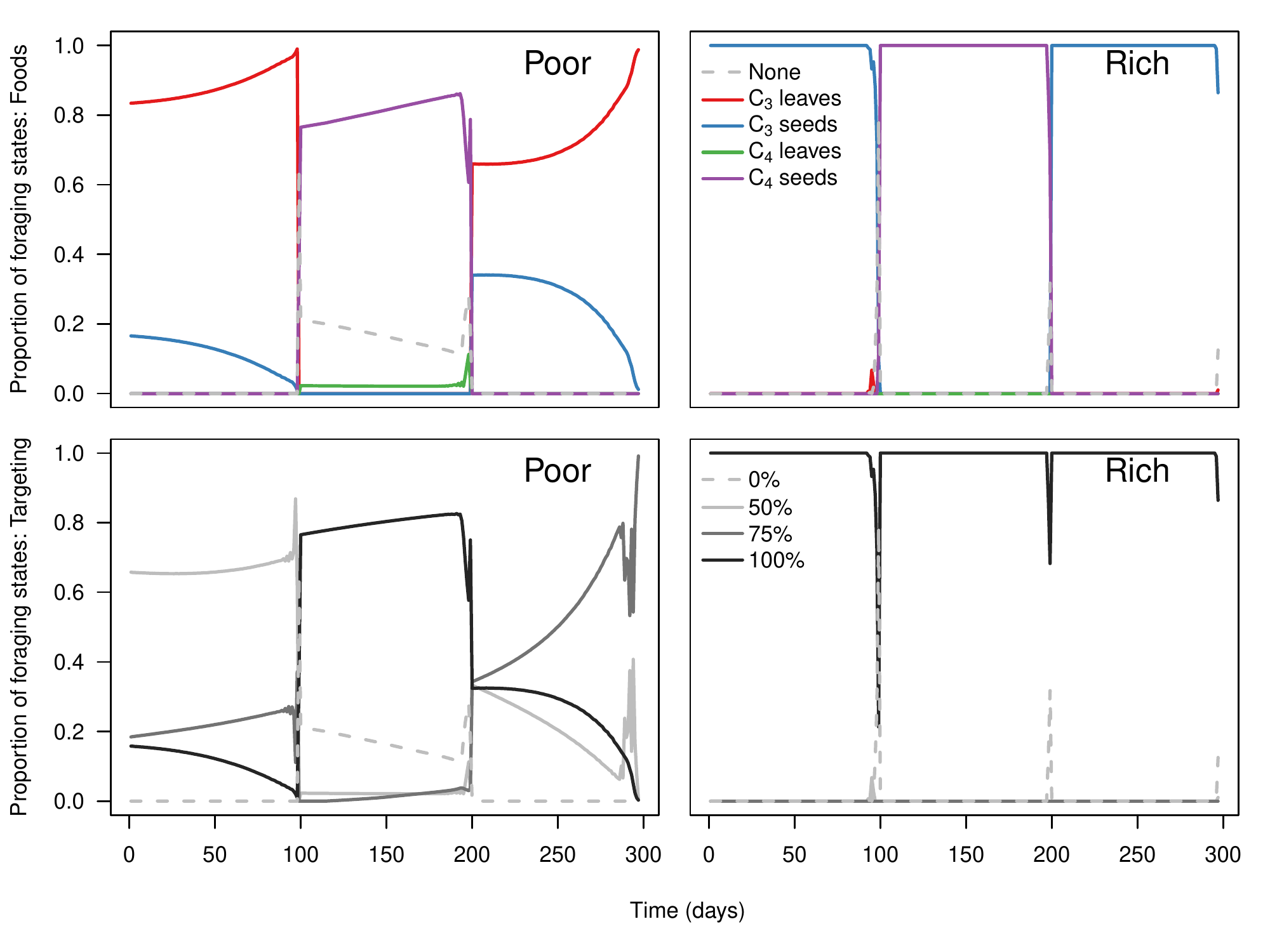}
\caption{
Top Row: Proportion of foraging states targeting different resources in poor (left) and rich (right) environments.
Targeting options include 1) none (gray), 2) \ct leaves (red), \ct seeds (blue), \cf leaves (green), \cf seeds (purple).
Bottom Row: Proportion of foraging states that employ different targeting weights in poor (left) and rich (right) environments.
Targeting weights range from none ($\tau=0$; dashed gray) to perfect targeting ($\tau=1$; black).  
}
\label{fig:foods}
\end{figure}

When the environment is poor and resources are difficult to find, more conservative strategies are promoted.
The strategy in which fallback resources are consumed and cached resources are saved (\emph{fallback and save}: F\&S) is only used when fat storage is very high, and by the avoidance of using cached resources  (fig. \ref{fig:forage}A,B).
Under these conditions, only fallback foods are targeted (fig. \ref{fig:foods}A).
Fallback foods (leaves) are more common but have lower energy density and are not cacheable, such that they can only replenish body reserves.
These qualities mean that such foods are not preferred, and in turn never heavily targeted (fig. \ref{fig:foods}B), allowing for the opportunistic incorporation of more preferred foods (seeds).
We observe this strategy during the pre-monsoon season in poor environments, where the consumer faces a year of survival in an unforgiving landscape.
The \emph{fallback and use} (F\&U) strategy is similar in that foraging for fallback foods is only carried out when consumer fat stores are high, but differs in that cached resources are used if available (figs. \ref{fig:forage}A,B).
We observe this strategy in poor environments during the post-monsoon season, where conserving cache savings carries little reward as the consumer approaches the time at which fitness is assessed (i.e. the terminal time $d_{\rm max}s_{\rm max}$).

When resources are more abundant, consumers can either take advantage of food abundance in the present, or prepare for future hard times.
In rich environments during the pre- and post-monsoon seasons, consumers adopt a \emph{replenish and use} (R\&U) strategy where the cache is used if it is available and replenished if it is not (figs. \ref{fig:forage}c,d).
In these cases, only cacheable seeds are targeted (fig. \ref{fig:foods}c,d).
Finally, monsoon conditions are characterized by not only greater densities of \cf foods but greater resource abundance overall, such that the risk of foraging and returning with an empty stomach is lower.
During the productive monsoon season in both rich and poor environments, consumers adopt a \emph{replenish and save} (R\&S) strategy where foraging occurs liberally unless fat stores are very low, and the cache is not used unless cache stores are very high.

Short-term hoarding strategies used by Carolina chickadees (\emph{Poecile carolinensis}; Paridae) demonstrated that foraging to replenish cache stores is expected to increase as consumer fat reserves decline; when the consumer is near starvation, effort is then redirected to restore body fat \cite{Lucas1991}.
Our model supports this expectation, but only in the monsoon season when cacheable foods (seeds) are common.
In the pre- and post-monsoonal environments, cacheable foods are rare, and this reverses the expected pattern.
During periods of resource limitation, we find that foraging effort increases only when the cache is low, or alternatively if consumer fat stores are high.
In these cases foraging effort is directed towards fallback rather than cacheable foods.

\begin{figure*}
\centering
\includegraphics[width=1\textwidth]{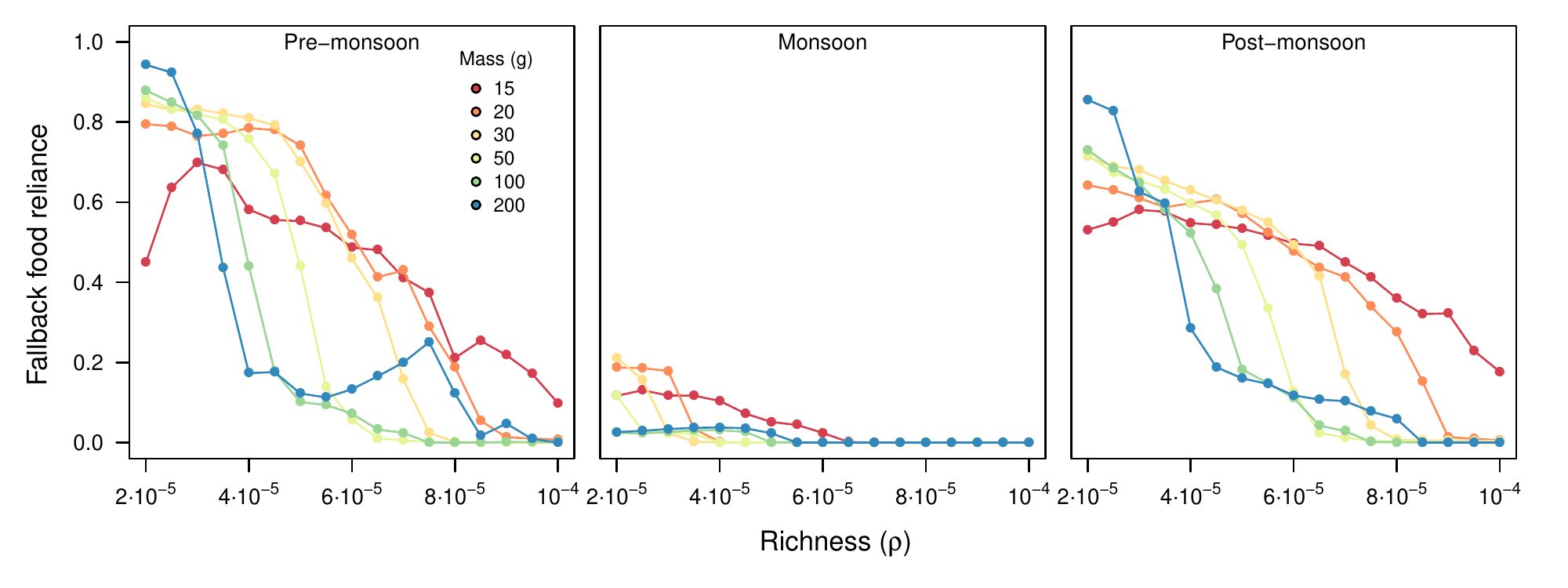}
\caption{
Proportional reliance on fallback foods (non-cacheable \ct and \cf leaves) as a function of environmental richness $\rho$, averaged over both fat and cache states as well as days for each pre- (left), full- (center), and post-monsoon (right) season.
}
\label{fig:fallback}
\end{figure*}

\subsection*{Body Size and the Nature of Fallback Foods} 
While fallback foods can appear to play a small role in diet, they may have a contrastingly large influence on the evolution of traits facilitating acquisition \cite{Ungar2004,Vogel2008,Marshall2009}, and may even have played a prominent role in human evolution \cite{Laden2005,Ungar2004,Yeakel2007,Constantino2009,Yeakel2013}.
However, a general theoretical understanding of how and when consumers utilize fallback foods is lacking.
At the Sevilleta, fallback foods such as \ct and \cf leaves are targeted to replenish fat stores when overall resource abundance is low.
Because the monsoon is characterized by overall higher resource abundance, fallback foods are of primary importance during the pre- and post-monsoon seasons.
Larger mammals store a larger percentage of their body weight as fat, such that fat mass scales superlinearly with body mass \cite{Lindstedt2002,Yeakel2018}.
As a consequence, the risks faced by mammals differs across body sizes, suggesting that the role of fallback foods may vary as well.

\begin{figure*}
\centering
\includegraphics[width=1\textwidth]{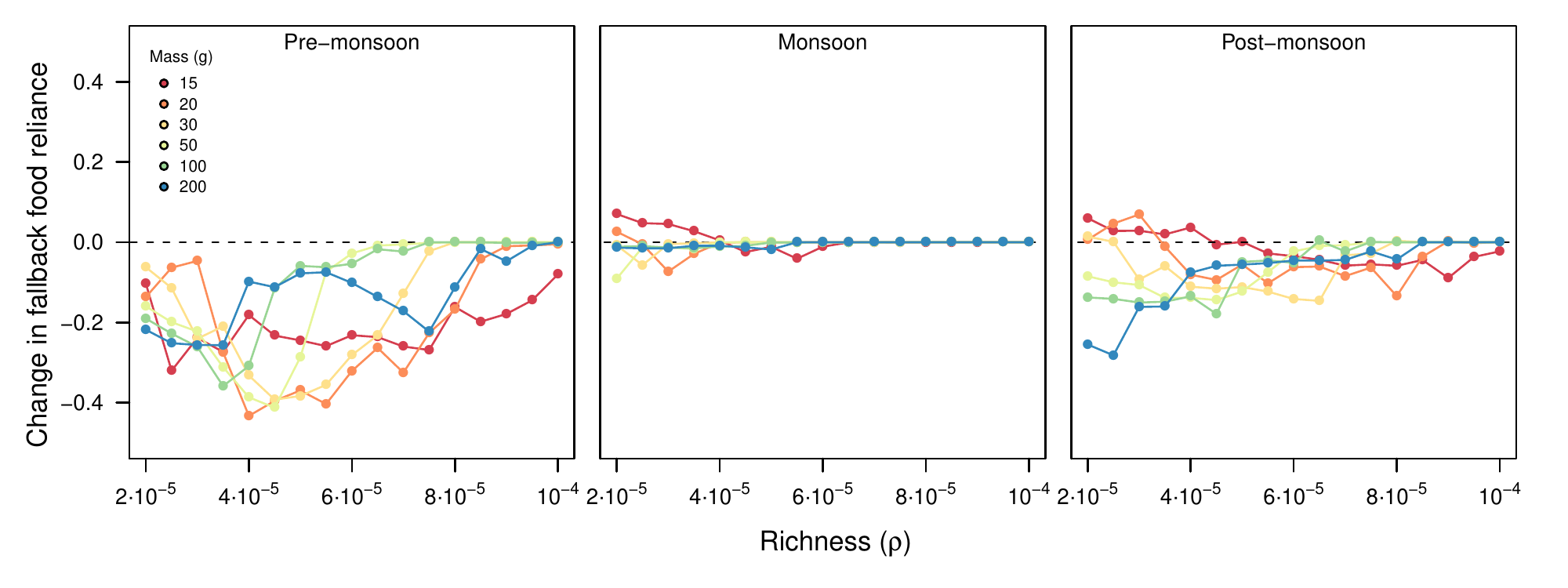}
\caption{
The difference in fallback food reliance (as measured in fig. \ref{fig:fallback}) between environments with uncertain seasonal transitions ($\sigma=20$) and those with deterministic seasonal transitions ($\sigma \approx 0$).
Reliance is averaged over both fat and cache states as well as days for each pre- (left), full- (center), and post-monsoon (right) season.
Positive values mean that consumer states in environments with deterministic seasonal transitions target more fallback foods; negative values mean that consumer states in environments with uncertain seasonal transitions target more fallback foods.
}
\label{fig:fallbackdiff}
\end{figure*}

Rodent species at the Sevilleta vary in body mass by over an order of magnitude, from $\sim 8-10$ g (e.g. \emph{Perognathus}) to $\sim 125-200$ g (e.g. \emph{Dipodomys} and \emph{Neotoma}).
Altering consumer body size influences a number of relationships in the foraging model, including the total amount of fat a consumer can store ($\propto M^{1.19}$), the metabolic costs associated with foraging and resting ($\propto M^{3/4}$), velocity ($\propto M^{0.21}$), as well as the area used to search for food ($\propto M^{0.75}$; table 2). %, and it stands to reason that our results should vary across this range.
We find that these consumers utilize fallback foods to different degrees during the pre-, full-, and post-monsoon seasons depending on body size and environmental quality, which we suggest has a number of important implications for the Sevilleta rodent community in the context of future environmental change.

% 1) Evolution of larger body size?
First, we find that reliance on fallback foods is nonlinear across consumer body size (fig. \ref{fig:fallback}).
Smaller consumers rely on fallback foods heavily, regardless of environmental quality, and this is due to the unaffordable risk of starvation associated with foraging for higher quality seeds.
However, this reliance is lessened for the smallest (15 g) consumer in the poorest environment, where it targets a larger proportion of higher quality seeds than expected.
That high-quality yet risky returns maximize fitness for consumers near starvation -- and small consumers are closer to starvation than larger consumers -- supports both theoretical and empirical expectations for risk-sensitive foraging \cite{Caraco1980,Craft2016}.
For all but the poorest environments, a greater reliance on fallback foods means that fewer energetic/cache states result in foraging behaviors that promote cache maintenance and replenishment.
Greater reliance on more ubiquitous foods lowers the uncertainty associated with maintaining adequate energetic reserves, and this becomes more important for smaller consumers with lower percent body fat.
Given that smaller consumers have a limited energetic tolerance, and cannot afford to conserve a substantial cache, it is likely that additional sources of uncertainty may disproportionately impact these species.

Intermediate to large consumers (30-200 g) tend to target preferred foods for a greater proportion of energetic states, but increase their reliance on fallback foods as environmental conditions deteriorate.
That smaller consumers rely more heavily on fallback foods over a larger range of environmental conditions is in general agreement with recent work showing that resource homogeneity provides greater fitness gains to smaller consumers, whereas larger consumers can more readily target resources that are harder to find without high fitness costs \cite{Bhat2019}.
In very poor environments, larger consumers that have a higher proportion of their body mass devoted to fat storage (e.g. \emph{Neotoma}) are expected to be less risk-sensitive.
In these cases, our model affirms general expectations that such consumers target the lower quality, more ubiquitous fallback foods.
Specifically to southwest desert ecosystems, this prediction is in line with observations of \emph{Neotoma} dependence on lower quality leaves \cite{Thompson1982,Kohl2014}.

% 3) Larger consumers should be expected to have larger caches
The rodent communities of the American Southwest have long inspired ecologists to ask: how do so many species with seemingly overlapping niches coexist in stochastic resource-limited environments \cite{Brown1975}?
We suggest that diverse use of fallback versus preferred foods among consumers of different body sizes, across the monsoon seasonal transition, may point to an important axis of differentiation that separates consumer niches.
While some species are expected to target \ct or \cf seeds, others are expected to target leaves or rely on their caches, and these proclivities vary with fat and cache state, in addition to body size and time of year.
When fallback foods are targeted, consumers prioritize their fat stores over cache maintenance and replenishment.
It follows that lowering reliance on fallback foods promotes a greater proportion of energetic states devoted to targeting seed resources, which by extension results in cache maintenance and larger cache sizes.
That larger species maintain larger caches is a well-known phenomenon in diverse rodent communities, and may promote coexistence of many species that share similar resources \cite{Price2000}.
Indeed, the two largest species at our field site, \emph{Dipodomys spectabilis} and \emph{Neotoma albigula}, are known to rely heavily on large caches \cite{Koontz2010}.

\subsection*{Uncertain Seasons}
We next examine how uncertainty in seasonal transitions, which is predicted to increase with climate change, impacts the expected foraging behaviors of rodent consumers.
We find that elevated uncertainty in the timing of seasonal transitions ($\sigma = 20$) has three important effects on the expected foraging strategies of rodent consumers.
First, as the uncertainty in predicting the resource landscape increases, alternative foraging strategies dominate during the most uncertain periods of the year, which in this case occur during the transitional periods between seasons (e.g. May-June).
Specifically, we find that foraging without targeting a specific resource maximizes fitness in poor environments once the uncertainty in the monsoonal onset begins (ca. day 80; appendix B), and remains an important strategy until uncertainty in the onset of the post-monsoon season ends (ca. day 220).
In rich environments, this foraging strategy is observed to maximize fitness only when seasonal uncertainty is highest (appendix B).
Foraging without a targeted resource means that the forager only scans the environment for the nearest resource, thereby minimizing travel costs without regard to resource quality or cacheability.
In environments without seasonal transition uncertainty, this strategy is rarely fitness maximizing (fig. \ref{fig:foods}), indicating that it is only worthwhile when resource distributions are unpredictable.

Second, seasonal transition uncertainty lowers the relative importance of fallback foods.
For consumers of all body sizes, the targeting of fallback foods averaged over energetic and cache states declines with increasing seasonal transition uncertainty (fig. \ref{fig:fallbackdiff}).
In the pre-monsoon, the decline in fallback food utilization is the most extreme, especially for smaller consumers in intermediate environments (fig. \ref{fig:fallbackdiff}a).
In the post-monsoon, the decline in fallback food utilization is greatest for larger consumers in poorer environments  (fig. \ref{fig:fallbackdiff}c).
This change in the use of fallback foods with increasing seasonal transition uncertainty means that more effort is spent foraging for cacheable resources that can be used to maintain larger caches.
Previous modeling efforts focused on daily timescales have shown that increasing uncertainty in foraging returns leads consumers to build fat rather than cache reserves \cite{Pravosudov:2001hb}. 
In contrast, our results show that longer-term uncertainty associated with seasonal transitions leads to strategies promoting cache maintenance.
That the largest redirection in effort is realized by smaller species during the pre-monsoon season suggests that it is these consumers most impacted by seasonal uncertainty, however as we show next this can be measured directly.

Third, seasonal transition uncertainty reduces consumer fitness, but the extent that fitness is reduced largely depends on both the body size of the consumer, and the size of its initial cache.
To measure the effect seasonal uncertainty has on expected fitness, we compare the probability of survival for a consumer foraging in an environment with deterministic seasonal transitions at the beginning of the year, $F(x,y,d=1,s=1|\sigma\approx0)$, to that of a consumer foraging in an environment with uncertain seasonal transitions, $F(x,y,d=1,s=1|\sigma=20)$.

\begin{figure}
\centering
\includegraphics[width=0.45\textwidth]{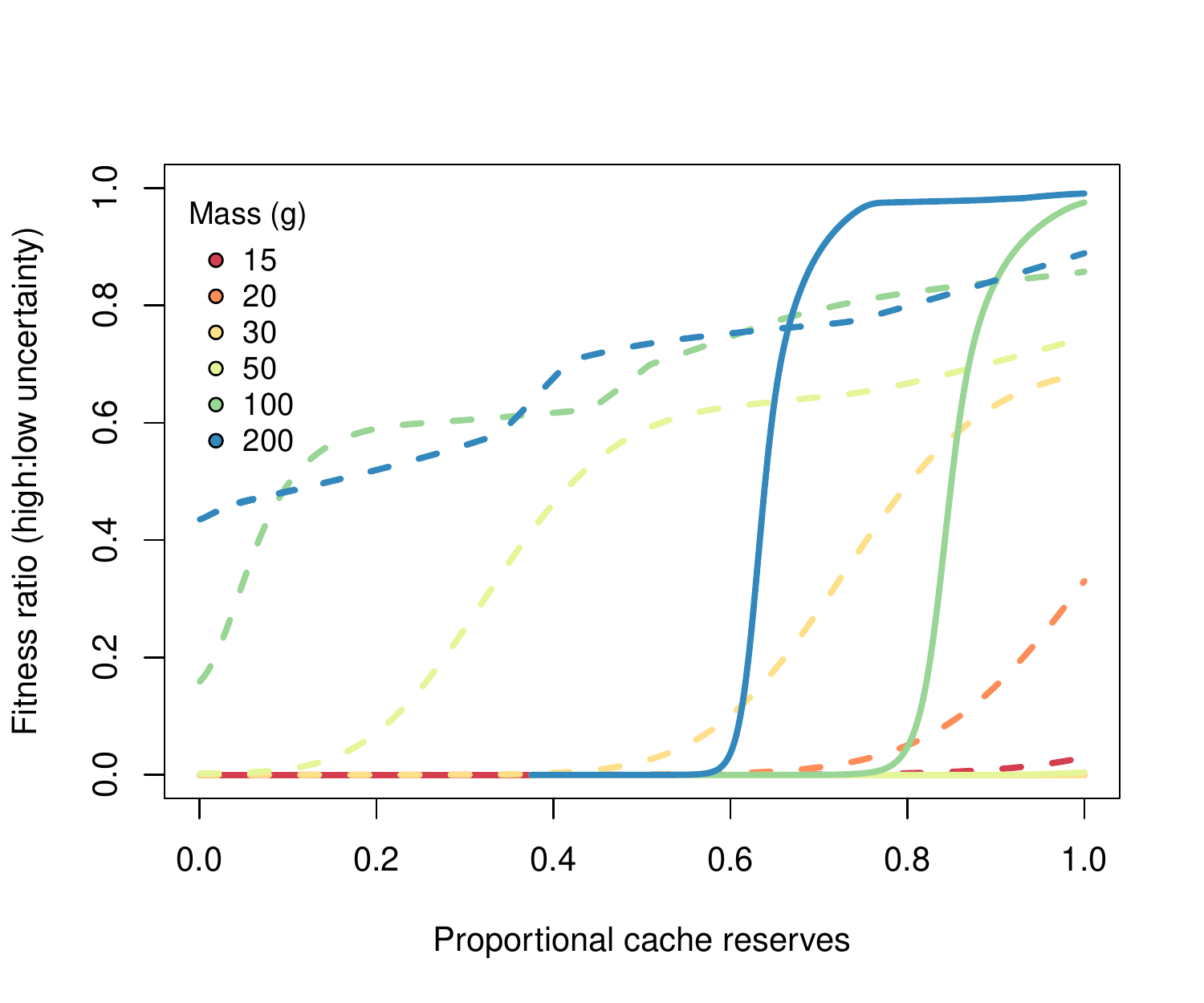}
\caption{
The ratio in fitness values for consumer states in environments with uncertain ($\sigma=20$) versus deterministic ($\sigma \approx 0$) seasonal transitions as a function of the initial size of the consumer's cache at the beginning of the year.
A value of 1 means that the uncertain and deterministic environments result in the same fitness values.
Values $<1$ mean that uncertain environments result in consumer fitness values lower than those in deterministic environments.
Solid lines: poor environments; dashed lines: rich environments.
}
\label{fig:onset}
\end{figure}

Our results reveal that while seasonal uncertainty lowers consumer fitness, a cache can compensate for these negative effects (fig. \ref{fig:onset}).
The cache size at which fitness gains compensate for seasonal transition uncertainty depends on both consumer body size and environmental quality.
In poor environments, the probability of survival is so low that when seasonal transitions are uncertain, compensation is accomplished only if the cache is $>60\%$ the maximum value, which is approximately equal to the amount needed to survive a single season (100 days) (solid lines, fig. \ref{fig:onset}).
Above this threshold, additional cache reserves have an exponential effect on fitness, fully compensating for the effects of seasonal uncertainty.
In contrast, the cache stores needed to compensate for the detrimental effects of seasonal uncertainty for consumers $<100$ g must be much larger than the maximum allowed in our framework.
Because such a cache is unrealistically large for these consumers, we suggest they will be most likely to suffer the fitness costs of seasonal uncertainty.

In rich environments, the cache threshold required to compensate for the negative effects of seasonal uncertainty is much lower (dashed lines, fig. \ref{fig:onset}).
In this case, the fitness advantage of a larger cache is more gradual, but again primarily benefits larger consumers, whereas consumers $<50$ g require close-to-maximum cache reserves for compensation.
Together, our findings suggest that while caches can compensate for seasonal transition uncertainty, they will provide a greater advantage to larger consumers.
While increased resource availability is expected to increase the range of body sizes over which the compensatory effects of a cache are realized, the smaller consumers remain most prone to the fitness costs of seasonal transition uncertainty.

\cite{Brodin1997} showed in a related model for willow tits (\emph{Poecile montanus}; Paridae) that even small amounts of cached resources are expected to have a large impact on survival during hard times.
Parids are scatter-hoarders, and store seeds in both short-term and long-term stores that require costly memorization and recall \cite{Brodin1992}.
While we focus here on the dynamics of central-place caching (larder hoarding), the rodents in the American Southwest practice both forms \cite{Schroder1979}, and which strategy is utilized may depend on the ability of a species to defend its cache \cite{Daly1992}.
Multiple models of hoarding behaviors for birds in the family Paridae have shown that during resource-depleted seasons, consumers redirect their efforts towards maintaining greater body fat stores \cite{Brodin1997,Clark2000}.
While the risks and demands associated with memorizing hidden caches are less important for central place foragers, our results support both of these findings.
Seasons that bring hard times require consumers to focus on body fat rather than cache maintenance, and it follows that those species that can endogenously store greater proportions of energy in the form of body fat are expected to have a survival advantage, which we show is magnified when seasonal transitions are uncertain.
Carrying more body fat may also entail higher predation risks, and predation is a major cause of mortality among desert rodents \cite{Sullivan2001}.
How such top-down risks integrate with those explored here will be the subject of future efforts.

Our model predictions point to a qualitative difference in the use of fallback foods and the fitness benefits of caching at the threshold body size of $\sim 50$ g.
Below this threshold fallback foods play a significant role even in richer environments, whereas above this threshold fallback foods are only utilized in very poor environments (fig. \ref{fig:fallback}).
Moreover, for consumers below this threshold caches must be very large -- equivalent to the energy needed to survive half a season -- to compensate for uncertain seasonal transitions.
Above this threshold, the fitness benefits are much larger for caches that are much smaller (fig. \ref{fig:onset}).
This prediction affirms empirical observations of cache use among rodents at our Sevilleta field site in the northern Chihuahuan Desert.
For example, nearly all of the four species of commonly caught food-caching granivorous Heteromyids are  $\geq 50$ g in body size, including three species of kangaroo rats (\emph{Dipodomys} spp.). The only exception is the small (8-10 g) silky pocket mouse (\emph{Perognathus flavus}). 
In contrast, the non-caching Cricetids that include members of the \emph{Peromyscus}, \emph{Onychomys}, and \emph{Reithrodontomys} genera are well below this threshold and range in size from $\sim 20–35$ g. 
Body condition data show that the Cricetids have body fat mass ranging from 8-12\% of their body weight, while the larger caching Heteromyids have body fat estimates of only 4–6\%. 
Interestingly, \emph{Perognathus} have body fat estimates (~8-10\%) that are similar to that of the slightly larger non-caching Cricetids, suggesting that the smallest species in the community use a combination of stored fat and cached reserves to persist in this uncertain resource landscape. 
Overall, our results support a size-dependent energetic basis for observed patterns in caching versus non-caching foraging strategies among members of the small mammal community in the northern Chihuahuan Desert.

Hard times are not only expected to become `harder' in a future defined by climate change, but more uncertain as well.
Consumers that make their living in ecosystems where transitions between seasons are severe may be more prone to the effects of these changes, particularly when caches are relied upon to counter the effects of uncertainty.
Here we have shown that consumers in a highly seasonal stochastic environment, characteristic of many temperate regions around the world, enact foraging strategies that utilize less available, nutrient-rich, and cacheable (preferred) foods differently than those that are more ubiquitous and non-cacheable (fallback), depending on consumer body size and environmental quality.
These different strategies reflect the increased risk of starvation faced by smaller consumers, where management of fat reserves often surpasses cache management, and these size-dependent effects are magnified by uncertain seasonal transitions.
Our framework suggests that it is the smaller-bodied consumers that benefit least from maintaining cached reserves, and are subject to the largest fitness detriments associated with seasonal uncertainty.
Moving into an uncertain future where hard times are the norm, there is much value in knowing which species are most prone to the fitness costs of additional uncertainty.
In the northern Chihuahuan Desert, it may be those species that can't save a cent.

\begin{table*}
  \begin{center}
\begin{tabular}{l l l l l l}
\hline
 & \ct leaves & \ct seeds & \cf leaves & \cf seeds & Units \\
\hline
Mean density $m_{\rm pre/post}$ & $9920$ & $1240$ & $620$ & $78$ & $\text{g~m}^{-2}$\\
Mean density $m_{\rm monsoon}$ & $4650$ & $581$ & $15500$ & $1938$ & $\text{g~m}^{-2}$\\
Energy density $g$ & $15$ & $21$ & $15$ & $21$ & $\text{kJ~g}^{-1}$\\
Digestibility $\epsilon$ & $33$ & $75^\dag$ & $25$ & $75^\dag$ & percent \\
\hline
\multicolumn{2}{l}{\footnotesize{${}^\dag$ Also the value of $\epsilon_{\rm cache}$}}\\
\end{tabular}
\caption{Resource characteristics.}
\end{center}
\label{tab:food}
\end{table*}

\begin{table*}
  \begin{center}
\begin{tabular}{l l l l}
\hline
Definition & Parameter & Units & Ref.\\
\hline
Consumer fat state & $X = x$ & kJ (var.) & \\
Consumer cache state & $Y = y$ & kJ (var.) & \\
Day in year & $d$ & n.a. & \\
Season in year & $s$ & n.a. & \\
\hline
Consumer mass & $M$ & $\text{g}$ & \\
Consumer fat mass & $0.02M^{1.19}+0.038{M^{1.0}}^\ast$ & $\text{g}$ & \citenum{Lindstedt2002}\\
Tissue energy density & $20$ & $\text{kJ}\cdot\text{g}^{-1}$ & \citenum{Stryer1995,Hou2008} \\
Consumer homerange & ${\propto M^{0.75}}^\dag$ & $\text{m}^2$ & \citenum{Calder1996,Lyons2019} \\
Consumer velocity & $v = 0.008 M^{0.21}$ & $\text{m~s}^{-1}$ & \\
Energetic gain & $K=k$ & kJ & \\
Resting metabolic rate & $b_{\rm basal} = 0.018 M^{3/4}$ & $\text{watt}\cdot\text{hrs}$ & \citenum{Brown2004}\\
Field metabolic rate & $b_{\rm field} = 0.047 M^{3/4}$ & $\text{watt}\cdot\text{hrs}$ & \citenum{Brown2004}\\
Time in bout & $t_{\rm bout} = 5$ & hrs & \\
Energetic cost (nf) & $c_{\rm nf} = b_{\rm basal}t_{\rm day}$ & kJ & \\
Energetic cost (f) & $c_{\rm f} = b_{\rm field}t_{\rm bout} + b_{\rm basal}(t_{\rm day}-t_{\rm bout})$ & kJ & \\
% Consumer fat mass & $c_{\rm f} = 0.047 M^{3/4}$ & g & \citenum{Lindstedt2002}\\
Stomach capacity & $x_{\rm stm} = 1.5 c_{\rm f}$ & kJ & \\
Cheek capacity & $x_{\rm chk} = 1.5 c_{\rm f}$ & kJ & \\
Richness & $\rho \in (2\cdot10^{-5},10^{-4})$ & n.a. & \\
Targeting weight & $\tau \in (0,1)$ & n.a. & \\ 

\hline
% \footnotesize{${}^*$ We assume 10\% of muscle mass can be consumed prior to starvation}
\multicolumn{2}{l}{\footnotesize{${}^\ast$ We assume 10\% of muscle mass can be consumed in addition to fat}}\\
\multicolumn{2}{l}{\footnotesize{${}^\dag$ intercept determined by richness $\rho$}}\\
\end{tabular}
\caption{Parameter definitions.}
\end{center}
\label{tab:params}
\end{table*}

\clearpage

% \bibliographystyle{amnatnat}
% \bibliography{zzforage}

\clearpage

\beginsupplement

\section*{Appendix A: Sensitivity to dispersion and home-range scaling}
\textbf{Resource dispersion} To consider central-place consumers of different body sizes that forage over home-ranges of different areas, we set the mean encounter rate to scale as $m_i = \rho \mu_i M^a$, where $\rho$ determines environmental richness across all potential resources, and the home-range scaling exponent is set to $a=0.75$.
Each resource $i$ is thus distributed according to a mean encounter rate $m_i$ and dispersion $\alpha_i$, where lower values of $\alpha$ correspond to patchier distributions.
We estimate the distribution of energetic returns numerically using a simulation where the distance from the consumer to a resource $i$ is drawn from an exponential distribution with mean $1/\Lambda_i$, where the random variable $\Lambda_i \sim {\rm Gamma}(\alpha_i,m_i/\alpha_i)$.
The expectation and variance of the distance from the consumer to the next resource $i$, $\Psi_i = \psi_i$ is thus
\begin{align}
  {\rm E}\{\Psi_i\} &= \frac{\alpha_i}{m_i(\alpha_i-1)}, \\ \nonumber
  {\rm Var}\{\Psi_i\} &= \frac{\alpha_i^3}{m_i^2(\alpha_i-1)^2(\alpha_i-2)}.
\end{align}
If the dispersion is relatively large, such that resources are spatially uniform, the effect of $\alpha_i$ on the mean and variance is minimal, which approximate as $1/m_i$ and $1/m_i^2$, respectively.
Because we are considering resources that are expected to have similar spatial clustering, we set $\alpha_i = 10$ for \ct and \cf leaves and seeds.

\begin{figure}[h!]
\centering
\includegraphics[width=0.5\textwidth]{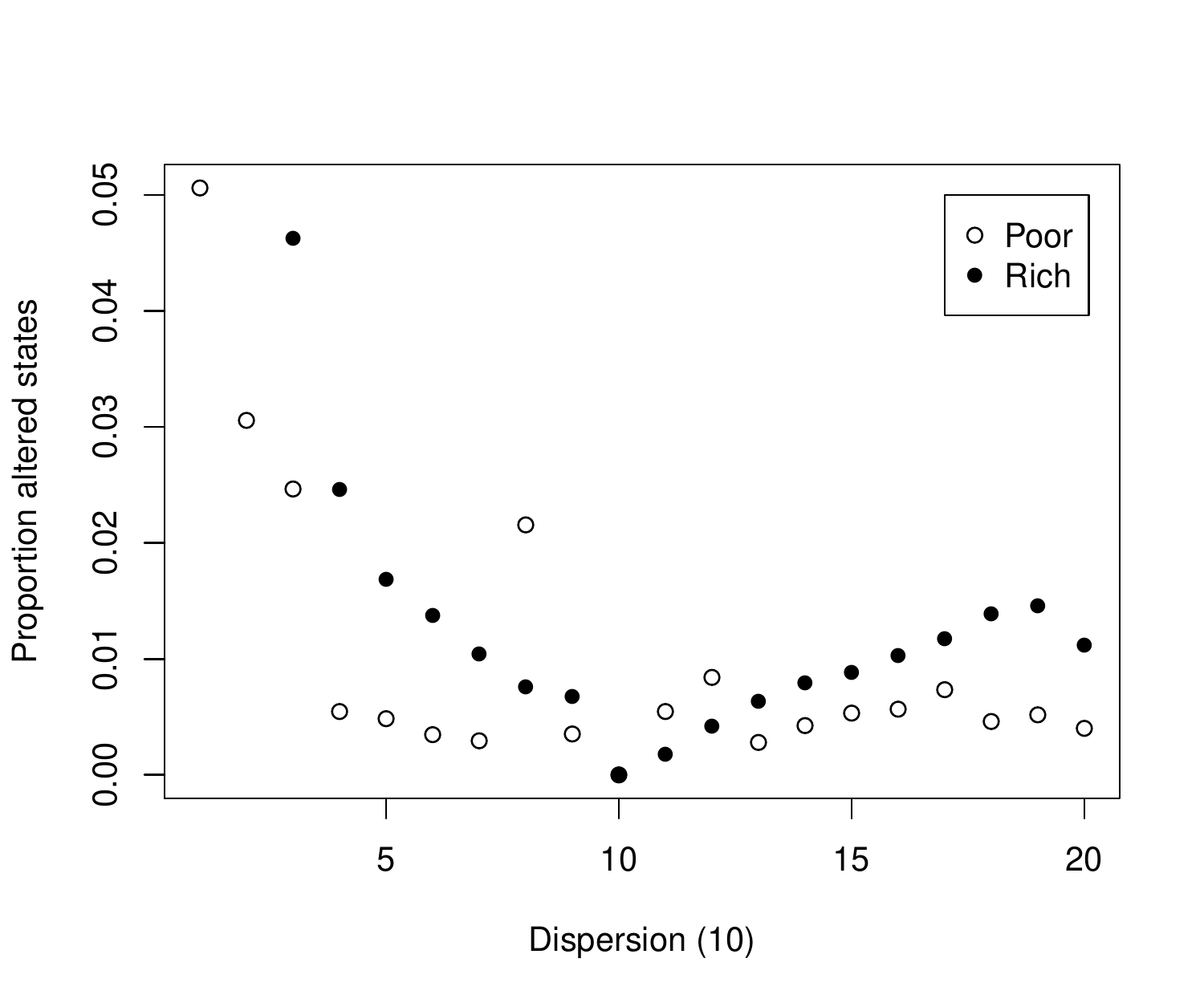}
\caption{
Sensitivity of analysis on the dispersion parameter $\alpha$. For the model described in the main text, we use dispersion values of 10 for all resources.
}
\label{fig:alpha}
\end{figure}

We examined the divergence of model results as a function of changing resource dispersion.
Model divergence was measured as the sum of differences in fitness-maximizing foraging strategies across all energetic $(X,Y)$ and temporal $(d,s)$ states.
We found that for values of $\alpha>10$, there was negligible divergence in model results, whereas as expected for very low values of $\alpha<10$, divergence increased (fig. \ref{fig:alpha}).

\textbf{Home-range scaling} Rodent consumers are central-place foragers with relatively small home-ranges that are spatially constrained by competition.
Because an individual consumer can cover the same ground multiple times within a foraging bout, resource availability is assumed to scale with home-range area, such that larger home-ranges promote increased resource availability and, by extension, higher effective densities and encounter rates.
A number of home-range scaling exponents $a$ have been measured \citep{Lindstedt1981,Calder1996}, ranging from sub- to super-linear.
For the model presented in the main text, we use a home-range scaling exponent $a=0.75$.

\begin{figure}[h!]
\centering
\includegraphics[width=0.5\textwidth]{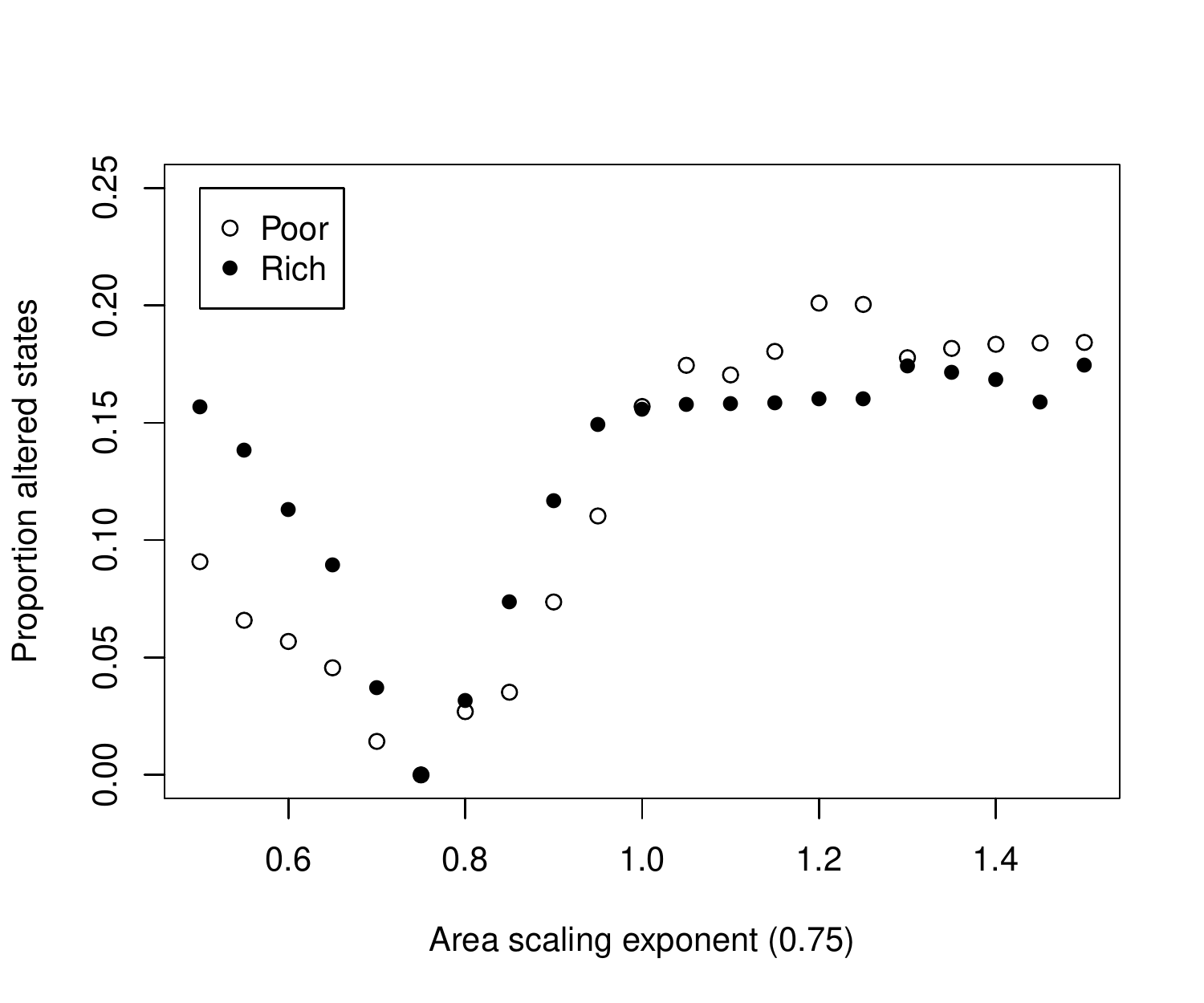}
\caption{
Sensitivity of analysis on the home-range scaling exponent. For the model described in the main text, we use a value of 3/4.
}
\label{fig:area}
\end{figure}

We examined the divergence of model results as a function of different scaling exponents, ranging from $a=0.5$ to $a=1.5$ (fig. \ref{fig:area}).
Smaller values of $a$ mean that larger consumers have only modest gains in resource availability, whereas larger values mean that larger consumers have much greater access to resources. 
For a consumer of $M=100$g, we find that changing the home-range scaling exponent has a significant, but predictable, influence on model divergence.
When the scaling exponent is small such that availability is lowered, there are fewer changes to fitness-maximizing strategies in poor landscapes because strategies are already optimized for low resource availability.
Under these conditions, rich environments become less rich, such that diverence is greater.
When $a$ is increased, resource availability is increased, and divergence is similar for both poor and rich landscapes.
Under these conditions, divergence plateaus after the home-range scaling exponent is above unity, as strategies for both poor and rich scenarios become optimized for very high resource availability landscapes.

\section*{Appendix B: Foraging in uncertain seasons}

\begin{figure}[h!]
\centering
\includegraphics[width=0.5\textwidth]{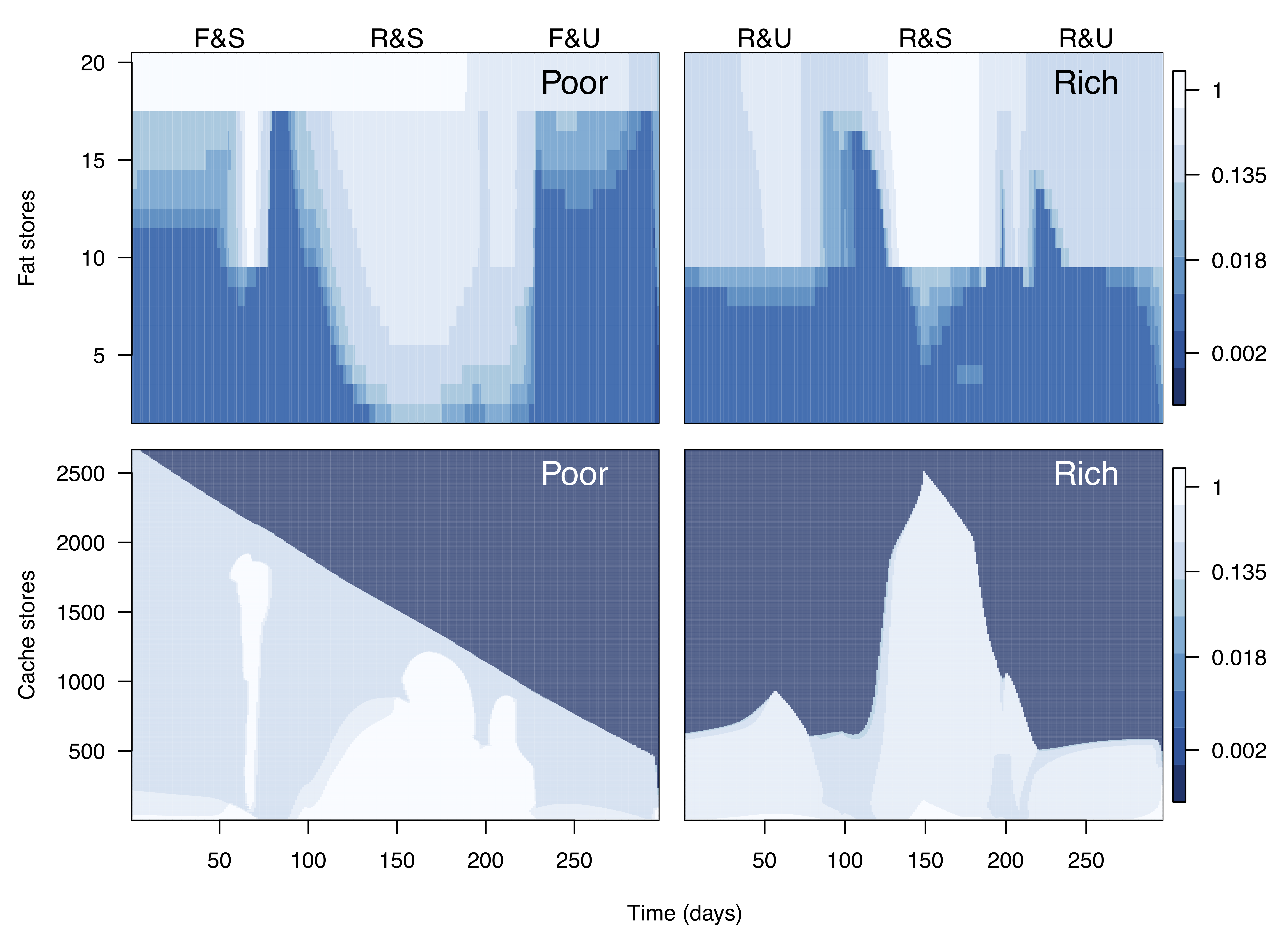}
\caption{
The proportion of states that result in active foraging versus not foraging and relying on cache reserves when seasonal transitions are uncertain ($\sigma=20$).
Values range from 100\% states that result in actively foraging (white) to 0\% (dark blue).
0\% active foraging implies 100\% relying on cached reserves.
To observe the dependence of foraging/not foraging strategies on consumer fat ($X$) and cache ($Y$) states as well as environmental richness, values are
(top-left) averaged across cache states in poor environments,
(bottom-left) averaged across fat states in poor environments,
(top-right) averaged across cache states in rich environments, and
(bottom-right) averaged across fat states in rich environments.
Note that the color scale is logged to emphasize smaller differences in consumer strategies.
}
\label{fig:forage20}
\end{figure}

Here we examine fitness maximizing foraging strategies when there is significant uncertainty in seasonal transitions ($\sigma=20$; see main text for details).
Seasonal uncertainty does not change the qualitative foraging strategies employed by consumers, but does impact behaviors around the transition period.
Foraging strategies can be summarized by combinations of two sets of alternative foraging modes: 
\emph{i}) \emph{fallback} versus \emph{replenish}: whether foraged foods consist of fallback resources (F), or those that can be used to replenish the cache (R), and
\emph{ii}) \emph{use} versus \emph{save}: whether cached resources are used (U) or saved (S) against future hardship.
In poor environments, we find that consumers actively forage across a larger range of fat storage states during both the uncertain transition between the pre- and full-monsoon season and the full- and post-monsoon season (fig. \ref{fig:forage20}).
This effectively expands the R\&S strategy that is emphasized during the monsoon season into the pre- and post-monsoon.
In rich environments, the reverse is true such that consumers tend to rely on their cache (if they have adequate reserves) during seasonal transitions.

\begin{figure}
\centering
\includegraphics[width=0.5\textwidth]{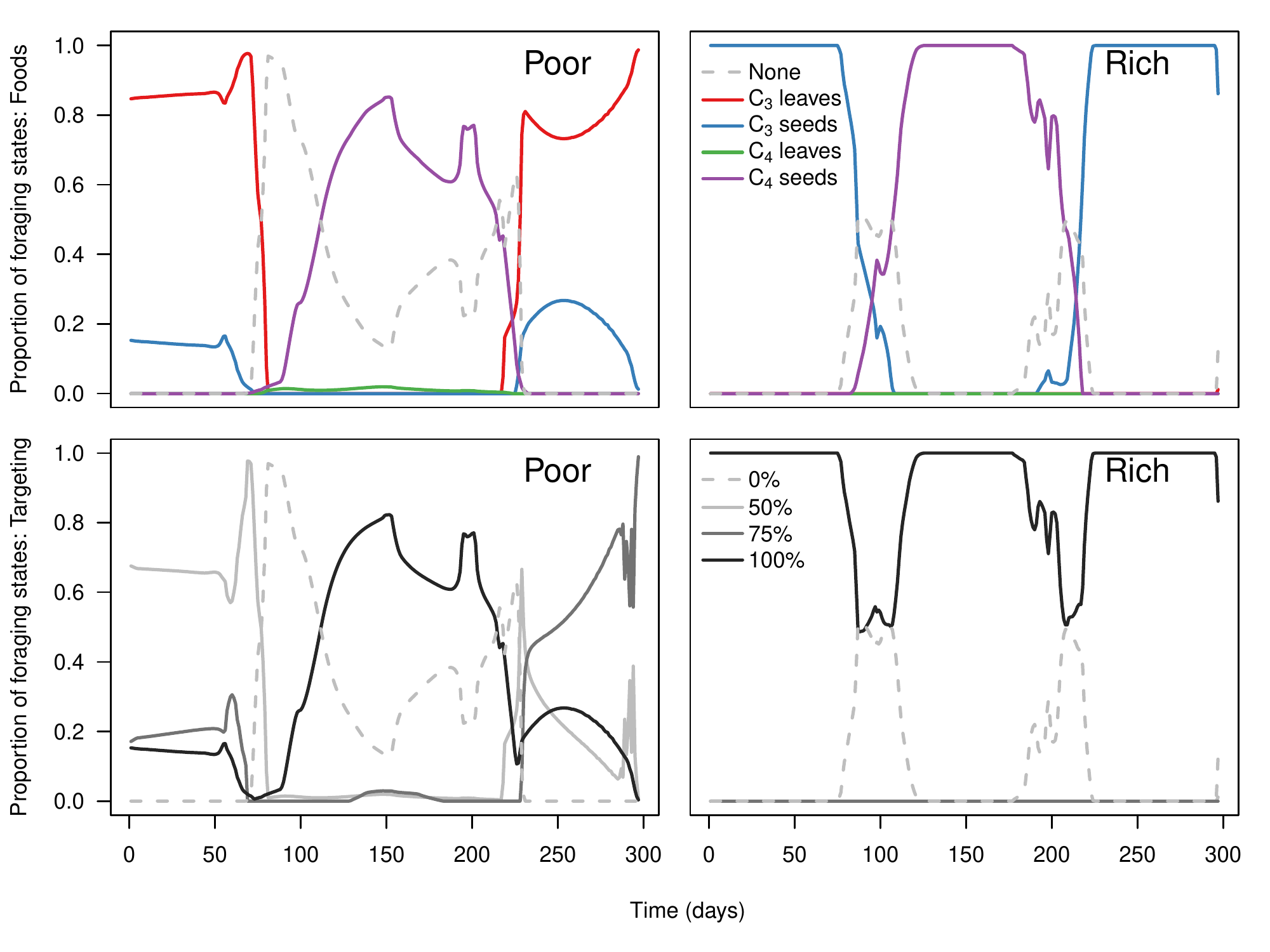}
\caption{
Top Row: Proportion of foraging states targeting different resources when seasonal transitions are uncertain ($\sigma=20$).
Targeting options include 1) none (gray), 2) \ct leaves (red), \ct seeds (blue), \cf leaves (green), \cf seeds (purple).
Bottom Row: Proportion of foraging states that employ different targeting weights when seasonal transitions are uncertain ($\sigma=20$).
Targeting weights range from none ($\tau=0$; dashed gray) to perfect targeting ($\tau=1$; black).}
\label{fig:foods20}
\end{figure}

By examining which foods consumers target when they do forage, we observe that seasonal uncertainty does not change the primary foods that are targeted during most of each season (fig. \ref{fig:foods20}).
However we do observe that during transition times, foraging without targeting any specific food plays a much larger role than when transitions are deterministic.
A non-targeting strategy means that the consumer always chooses the closest resource as it forages.
When resource distributions are unpredictable, choosing the closest food conserves time and prioritizes fat reserves, as cacheable foods (seeds) are less common than non-cacheable foods (leaves).

\end{document}